# State of the Art Survey of Deep Learning and Machine Learning Models for Smart Cities and Urban Sustainability

Saeed Nosratabadi[1], Amir Mosavi[2,3], Ramin Keivani[2],

Sina Ardabili[4], and Farshid Aram[5]

1 Institute of Business Studies, Szent Istvan University, Godollo 2100, Hungary
2 School of the Built Environment, Oxford Brookes University, Oxford, UK amir.mosavi@kvk.uni-obuda.hu
3 Kalman Kando Faculty of Electrical Engineering, Obuda University, Budapest, Hungary
4 Institute of Advanced Studies Koszeg, Koszeg, Hungary  5 Escuela Técnica Superior de Arquitectura, Universidad Politécnica de Madrid-UPM, 28040 Madrid, Spain

**Abstract.**

Deep learning (DL) and machine learning (ML) methods have recently contributed to the advancement of models in the various aspects of prediction, planning, and uncertainty analysis of smart cities and urban development. This paper presents the state of the art of DL and ML methods used in this realm. Through a novel taxonomy, the advances in model development and new application domains in urban sustainability and smart cities are presented. Findings reveal that five DL and ML methods have been most applied to address the different aspects of smart cities. These are artificial neural networks; support vector machines; decision trees; ensembles, Bayesians, hybrids, and neuro-fuzzy; and deep learning. It is also disclosed that energy, health, and urban transport are the main domains of smart cities that DL and ML methods contributed in to address their problems.

Keywords: Deep learning, Machine, learning, Smart cities, Urban sustainability, Cities of future, Internet of things (IoT), Data science, Big data

## 1 Introduction

Global urbanization is growing at a fast pace [1]. In the near future, major population of the world will be moving to the cities [2]. This trend will be extremely challenging for the land use management, sustainable urban development, food supply, safety, security, and human well-being in general [3, 4].

The emerging technologies and novel concepts for smart cities have been very promising to encourage a brighter future in dealing with the cities of the future. The artificial intelligence applications such as internet of things (IoT) [5], machine learning (ML) [6–9], deep learning (DL) [10] and big data [7, 11–15], have been essential in supporting the smart cities evolution and technological advancement. Among them,

ML methods have been contributing to various application domains with promising results in, e.g., mobility management and monitoring, city planning, resource allocation, energy demand and consumption prediction, food supply and production prediction, air pollution monitoring and prediction, etc. [16–21].

Literature includes an adequate number of state of the art review papers and comparative analysis on the general applications of ML and DL methods [22–35]. The trends of the advancement of ML and DL methods are reported to be hybrid and ensemble methods [36–46]. Considering the smart cities research, although, there have been various surveys on the applications of artificial intelligence, ML and DL methods, an insight into the popular methods, classification of the methods, and future trend in the advancement of novel methods are not given yet [47–54]. Thus, the current research aims to fill this gap through providing the state-of-the-art of ML and DL methods used for smart cities toward a more sustainable approach for urban sustainability. To do so, a novel classification is used to identify the most popular ML and DL models and review them in individual groups according to the methods used. This paper further contributes to identifying future trends in the advancement of learning algorithms for smart cities. Unlike in other fields, e.g., atmospheric sciences and hydrology were hybrids, and ensemble ML models have increased in popularity, in the smart city domain DL applications are dominant.

## 2 ML and DL Models for Smart Cities

To identify the most relevant literature in the realm of using ML and DL methods for smart cities and sustainable urban development we explored the web of science (WoS) and Scopus with the following search keywords: "smart cities" or "sustainable urban development" and all the existing ML and DL methods [12]. Figure 1 represents the overall research results showing the exponential growth in using the ML and DL methods for smart cities and sustainable urban development. The search in the major research databases revealed that the popularity of machine learning is still limited in only a handful of ML and DL methods.

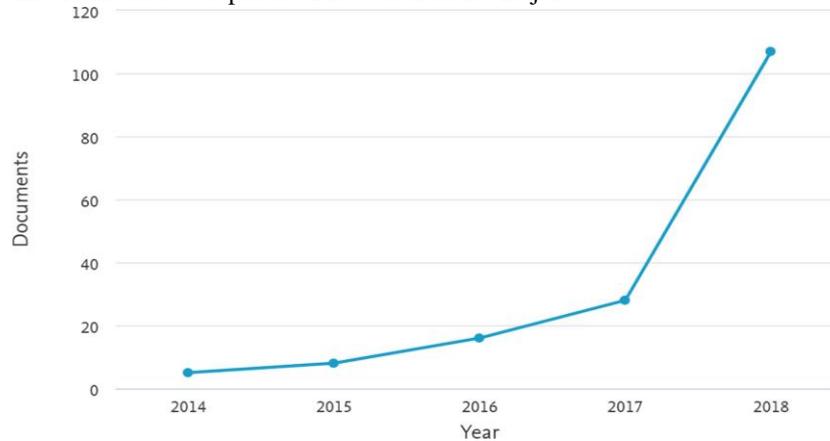

Fig. 1. The exponential growth in using ML and DL methods for smart cities and sustainable urban development (source: web of science)

In the following the popular ML and DL methods are identified and reviewed in the classified tables based on the methods used.

### 2.1 Artificial Neural Networks in Smart Cities

As it is summarized in Table 1, several research papers have applied artificial neural networks (ANNs) in the context of smart cities. ANNs have many applications in smart cities, including hazard detection, water supply, energy, and urban transport. Ullah et al. [55], for instance, develop infrastructures for smart lightning detection system aided by ANNs. Yuan et al. [56] provide an approach to improve the stability of wind turbines in smart cities. Rojek and Studzinski [57] design a method to detect water leaks in smart cities utilizing neural networks. Pan et al. [58] and Vlahogianni et al. [59] have applied ANNs to provide solutions for urban transport in smart cities.

Table 1. Notable applications and contributions of ANNs in the smart cities

| Literature references | Contributions | Application domains |
|---|---|---|
| Ullah et al. [55] | Utilizing artificial neural network to develop Infrastructures for the smart lightning detection system | Hazard detection |
| Yuan et al. [56] | Increasing the stability of wind turbines in smart cities | Energy |
| Rojek and Studzinski [57] | Providing a solution for water leaks detection utilizing neural networks in smart cities | Water supply and energy |
| Pan et al. [58] | Network traffic prediction for the smart cities using DE-BP neural network | Urban transport |
| Vlahogianni et al. [59] | Developing a prediction system for real-time parking in smart cities | Urban transport |

### 2.2 Support Vector Machines

Support vector machines (SVMs) is another machine learning method that can be used to deal with smart cities' problems. SVMs have been applied in different aspects of a smart city such as water supply, Energy, evaluation and management of smart, and health domains. Table 2 elaborates the contribution and domain application of SVMs in the smart cities, where Livingston et al. [60] provide solutions to improve water utilization in smart cities utilizing SVM. Chen and Zhang [61] providing a model to evaluate green smart cities in China. Chui et al. [62] propose an innovative approach to evaluate energy sustainability in smart cities. Ultimately, Aborokbah et al. [63] and Muhammad et al. [64] utilize SVMs to provide solutions for the health industry of smart cities.

Table 2. Notable applications and contributions of SVMs in the smart cities

| Literature references | Contributions | Application domains |
|---|---|---|
| Livingston et al. [60] | Improving water utilization in smart cities using SVM | Water supply |
| Chen and Zhang [61] | Providing a model to assess green smart cities in China | Evaluation and management of smart cities |
| Chui et al. [62] | Proposing a novel approach to evaluate energy sustainability in smart cities | Energy |
| Aborokbah et al. [63] | Designing an adaptive context-aware decision computing paradigm for intensive health care delivery in smart cities | Health |
| Muhammad et al. [64] | Designing a facial-expression monitoring system to improve healthcare in smart cities | Health |

2.3   Tree-Based Models (Decision Trees)

Another ML method which has been applied to solve problems of different aspects of smart cities is Decision trees (DTs) method. As it is detailed in Table 3, researchers have applied DTs to address the issues related to businesses, air pollution, urban transport, and food to develop a smart city. Ilapakurti et al. [65] formulate adaptive edge analytics for creating a memorable customer experience and venue brand engagement for smart cities. Orlowski et al. [66] design business models to measure air quality for smart cities for using IoT (Internet of Things) and SaaS (software as a service). Mei et al. [67] design an incentive framework for an intelligent traffic system based on initiative game-theory. Vuppalapati et al. [68] design a smart dairy model using IoT sensor network measuring cattle's health issues, milk production prediction, and productivity improvement.

Table 3. Notable applications and contributions of Decision trees in the smart cities

| Literature references | Contributions | Application domains |
|---|---|---|
| Ilapakurti et al. [65] | Formulating adaptive edge analytics for creating a memorable customer experience and venue brand engagement for smart cities | Businesses |
| Orlowski et al. [66] | Designing business models to measure air quality for smart cities for using IoT (Internet of Things) and SaaS (Software as a Service) | Air Pollution |
| Mei et al. [67] | Designing an incentive framework for an intelligent traffic system based on initiative game-theory | Urban transport |
| Vuppalapati et al. [68] | Designing a smart dairy model using IoT sensor network measuring cattle's health issues, milk production prediction and productivity improvement | Food |

2.4   Ensembles, Bayesian, Hybrids, and Neuro-Fuzzy

In addition to ANNs, SVMs, and DTs, which have a remarkable contribution to smart cities; ensembles, Bayesian, hybrids, and neuro-fuzzy have been applied to address the issues in the domains such as energy, urban governance, evaluation and management of smart cities, and health for smart cities. Table 4 summarizes the articles that have used ensembles, Bayesian, hybrids, and neuro-fuzzy to deal with different problems in smart cities. Where Nguyen et al. [69] design a sustainable model for urban landscape evolution city. Taveres-Cachat et al. [70] propose a framework to build a zero-emission neighborhood using responsive building envelope. Ju et al. [71] design a framework to apply citizen-centered big data for governance intelligence in smart cities. Tan et al. [72] develop an adaptive neuro-fuzzy inference system approach for urban sustainability assessment. Finally, Sajjad et al. [73] provide a quality computer-aided blood analysis system to discover and count the white blood cells in blood samples. Their approach contributes to making the healthcare industry smart in the smart city (see Table 4).

Table 4. Notable applications and contributions of ensembles, Bayesian, hybrids, and neurofuzzy in the smart cities

| Literature references | Contributions | Application domains |
| --- | --- | --- |
| Nguyen et al. [69] | Designing a sustainable model for urban landscape evolution city | Evaluation and management of smart cities |
| Taveres-Cachat et al. [70] | Designing a framework to build a zeroemission neighborhood using responsive building envelope | Energy |
| Ju et al. [71] | Proposing a framework to apply citizencentered big data for governance intelligence in smart cities | Evaluation and management of smart cities |
| Tan et al. [72] | Developing an adaptive neuro-fuzzy inference system approach for urban sustainability assessment | Evaluation and management of smart cities |
| Sajjad et al. [73] | Providing a quality computer-aided blood analysis system to the discover and count the white blood cells in blood samples | Health |

## 2.5 Deep Learning

Deep learning methods have had numerous and various applications in developing smart cities. The body of research, aided by such methods, has contributed to different aspects of a smart city such as energy sector, health, transportation, and even management of smart cities. As it is presented in Table 5, Luo et al. [74] and VázquezCanteli et al. [75] utilize deep learning to provide solutions in the energy sector for the smart cities. Where Luo et al. [74] design a system for a short-term energy prediction for a smart city. Vázquez-Canteli et al. [75] develop an integrated simulation environment to manage energy intelligently. Baba et al. [76] provide a sensor network for violence detection in smart cities. Reddy and Mehta [77] propose a system for smart traffic management for smart cities. Muhammed et al. [48] and Obinikpo and Kantarci [78] applied deep learning to deal with the concerns in the health sector. Finally, Madu et al. [79] propose a framework to evaluate urban sustainability utilizing deep learning.

Table 5. Notable applications and contributions of deep learning in the smart cities

| Literature references | Contributions | Application domains |
| --- | --- | --- |
| Luo et al. [74] | Crafting a system for a short-term energy prediction for smart city | Energy |
| Baba et al. [76] | Designing a sensor network for violence detection in smart cities | Security |
| Reddy and Mehta [77] | Proposing a system for Smart traffic management in smart cities utilizing reinforcement learning algorithm | Urban transport |
| Vázquez-Canteli et al. [75] | Developing an integrated simulation environment to manage energy intelligently in smart cities | Energy |

| Muhammed et al. [48] | Providing a ubiquitous healthcare framework, utilizing edge computing, deep learning, big data, high-performance computing (HPC), and the Internet of Things (IoT) | Health |
|---|---|---|
| Obinikpo and Kantarci [78] | Providing all the applications of deep learning methods used in sensed data for prediction in smart health services | Health |
| Madu et al. [79] | Providing a framework to evaluate urban sustainability utilizing deep learning | Evaluation and management of smart cities |

## 3 Discussion and Conclusions

The current paper provided a comprehensive state-of-the-art of ML and DL methods used for smart cities toward urban sustainability. A novel classification is used to identify the most popular ML and DL model and review them in individual groups according to the methods used. The ANNs, SVMs, DTs, Ensembles, Bayesians, and neuro-fuzzy methods have been seen as the most used machine learning methods. The finding revealed that ML and DL methods had remarkable contributions to the development of smart cities. It is also shown that energy, health, urban transport, evaluation and management of smart cities, water supply, businesses, air pollution, food, urban governance, security, and hazard detection are different domains of smart cities have borrowed ML and DL methods to deal with the similar problems.

Energy was the sector which has most leveraged the ML and DL methods in development of smart cities as four out of the 5 methods concerned in this study (artificial neural networks, support vector machines (SVMs), Ensembles, Bayesians, hybrids, and neuro-fuzzy, deep learning) are applied to provide different solutions for this sector. After energy, health, urban transport, and evaluation and management of smart cities are the other smart cities domains that have had most attention by the researchers in the standard fields of ML and DL methods and smart cities where at least three different ML and DL methods are applied to address their research questions. This paper also reveals an unexpected result, i.e., the immense popularity of DL methods. The DL methods have been seen dramatically popular in smart city applications mainly published in 2018 and 2019. This paper further identified future trends in the advancement of learning algorithms for smart cities. The trend in smart cities have shown to follow the trend in the overall trend which is a shift toward the advancement of the more sophisticated hybrid, ensemble and deep learning models, as also shown in [80–88].